\documentclass[a4paper]{article}

\usepackage[utf8]{inputenc}
\usepackage[T1]{fontenc}

\usepackage[a4paper,margin=2.5cm]{geometry}

\usepackage{lmodern}

\usepackage{amsmath, amssymb}
\usepackage{tensor}
\usepackage[normalem]{ulem}

\usepackage{cite, aas_macros}
\usepackage[sort&compress,numbers]{natbib}

\usepackage[dvipsnames]{xcolor}
\usepackage{url}
\usepackage[colorlinks=true, citecolor=Blue, linkcolor=Red]{hyperref}

\newcommand{\matM}{\blue{\mathfrak{m}}}

\DeclareMathOperator{\divergence}{div}
\newcommand{\cref}[1]{\red{\ref[1]}}

\let\t\tensor
\let\p\partial

\newcommand{\redsigma}{\sigma}
\newcommand{\redtau}{\tau}

\newcommand{\ptcheck}[1]{\ptc{checked on #1}}

\newcommand{\reda}[1]{\ptcr{change or addition or rewording}\color{red}}

\newcommand{\blue}[1]{{\color{Blue}#1}}
\newcommand{\red}[1]{{\color{red}#1}}

\newcounter{mnotecount}[section]
\renewcommand{\themnotecount}{\thesection.\arabic{mnotecount}}

\newcommand{\mnote}[1]{%
\protect{\stepcounter{mnotecount}}%
\textsuperscript{$\bullet$\themnotecount}%
\marginpar{\raggedright\tiny\em$\hspace{-1em}\bullet$\themnotecount: #1}%
}

\newcommand{\ptc}[1]{\mnote{{\bf ptc:}#1}}
\newcommand{\ptcr}[1]{{\color{red}\mnote{{\color{red}{\bf ptc:}#1} }}}

\def\R{\mathbb R}

\usepackage{textgreek}
\def\permeability{\text{\textmu}}
\def\permittivity{\text{\textepsilon}}

\renewcommand{\red}[1]{#1}
\renewcommand{\blue}[1]{#1}

\usepackage{authblk}

\renewcommand{\ptcheck}[1]{}

\begin{document}
\title{No Proca Photons}

\author[1,2]{F.~Steininger}
\author[1,2]{T.~B.~Mieling}
\author[2]{P.~T.~Chruściel}
\affil[1]{\footnotesize University of Vienna, Faculty of Physics, Vienna Doctoral School in Physics, Boltzmanngasse 5, 1090 Vienna, Austria}
\affil[2]{\footnotesize University of Vienna, Faculty of Physics and Research platform TURIS, Boltzmanngasse 5, 1090 Vienna, Austria}

\date{\today}
\maketitle

\begin{abstract}
We show that the Proca equation in vacuum, as well as its plausible modifications in dielectric media, is incompatible with experimental evidence, no matter how small the Proca mass is.
\end{abstract}


\section{Introduction}

Both Maxwell’s equations and the standard model of particle physics assume photons to be massless.
The question whether this is so remains intriguing, as mechanisms have been described \cite{BONETTI2017203,AdelbergerDvaliGruzinov}
which provide photons with an effective mass.
However,
a non-zero photon mass would imply corrections to Coulomb’s inverse-square law, and would challenge the current picture of photons being gauge bosons, see, e.g., \cite{GoldhaberNietoRMP,Tu,BONETTI2017203} and references therein.

The standard modification of the microscopic Maxwell equations to include a photon mass $\mu > 0$ is given by Proca’s equation
\begin{align}
	\label{eq:proca j vacuum}
	&\t\eta{^\alpha^\beta} \t\p{_\alpha} \t F{_\beta_\gamma}
		- \frac{\mu^2 c^2}{\hbar ^2} \t A{_\gamma}
		= - \eta_{\gamma\alpha} \t j{^\alpha}
	\,,
	&
	&\text{with}&
	&
	\t F{_\alpha_\beta}
		= \t\p{_\alpha} \t A{_\beta} - \t\p{_\beta} \t A{_\alpha} \,,
\end{align}
where $\t\eta{_\alpha_\beta} = \operatorname{diag}(-1,+1,+1,+1)$ is the Minkowski metric and $\t j{^\alpha}$ the electric current, which is assumed to satisfy the continuity equation,  $\t\p{_\alpha} \t j{^\alpha} = 0$. Henceforward, we will set $\hbar$ and $c$ equal to $1$.

Experiments testing this equation have produced more and more stringent upper bounds on the photon mass $\mu$, with the current Particle Data Group bound~\cite{ParticleDataGroup:2022pth} being $\mu < 10^{-18} \,\text{eV}/\text{c}^2$; compare~\cite{AdelbergerDvaliGruzinov} for a discussion of more stringent bounds based on astrophysical considerations.

While it is generally assumed that a small photon mass $\mu$ in \eqref{eq:proca j vacuum} produces only small deviations from solutions to Maxwell’s equations \cite{GoldhaberNietoRMP}, we show that this is not the case.
In particular, we show that \eqref{eq:proca j vacuum} implies matching conditions at material interfaces which deviate significantly from those of Maxwell’s equations. Considering coaxial cables as an example, we show the following result:

\emph{Assuming charge conservation, and that the field strength contains no Dirac-delta distributions, the Proca equation \eqref{eq:proca j vacuum} in a coaxial cable consisting of vacuum enclosed by ideal conductors rules out the existence of transverse electromagnetic (TEM) modes, no matter how small $\mu$ is.}

The condition that $\t F{_\alpha_\beta}$ contains no Dirac-delta distributions is motivated by the requirement of a well-defined energy-momentum tensor, which would otherwise contain squares of Dirac-deltas.

We further demonstrate that the same result remains valid for two natural generalizations of equation \eqref{eq:proca j vacuum} to linear isotropic dielectrics.

The argument proceeds by showing that $\mu \neq 0$ in \eqref{eq:proca j vacuum} implies an inequality, which forbids TEM modes, between the propagation constant of the electromagnetic field and its frequency.
It should be kept in mind  that the TEM modes are the main propagation modes in coaxial cables.

This leaves one with  two options of either discarding Proca’s equation, or calling into question the assumptions leading to these results.
While we indicate below a possible modification of this equation which restores TEM modes at the cost of introducing a preferred vector field associated with the symmetry axis of the coaxial cable, no such modification is available in the absence of spatial symmetry.

\section{Proca Modes in Coaxial Cables}
\label{sec:vacuumProca}

Recall that coaxial cables are composed of a solid copper core, surrounded by a dielectric insulator, which is in turn surrounded by woven copper shielding, and covered by a plastic sheath.
Copper is modelled as a perfect conductor, meaning that the electromagnetic field has to vanish within the metal.
The field is confined between the conductors, hence no field is present in the sheath and ambient air.
The dielectric’s refractive index is often close to unity, so that the region between the core and shielding can be modelled as a vacuum as far as the electromagnetic properties are concerned. We also make this assumption in this section, but allow for non-trivial refractive indices in the next section.

To proceed, we start by determining the boundary conditions satisfied by the fields.
By definition, the field $\t F{_\alpha_\beta}$ vanishes inside ideal conductors. There is an induced current $\t j{^\alpha}$ which is non-zero only on the interfaces.
Thus, by \eqref{eq:proca j vacuum}, $\t A{_\alpha}$ vanishes within the conductors as well.

Let us show that this, together with our hypotheses listed above, implies continuity of $\t A{_\alpha}$ at the interface.
For this, let us denote by $U \times \R \subset \R^3$ the vacuum region between the conductors, where $U \subset \R^2$ is an annulus with interior radius $r_1$ and exterior radius $r_2$, and let $\p U $ be the boundary of $U$.
In cylindrical coordinates ($ t , r , \theta , z $), the requirement that $\t F{_\alpha_\beta} $ contains no Dirac-delta distributions immediately implies continuity of $\t A{_t}$, $\t A{_\theta}$ and $\t A{_z}$ at $\p U$.

To show continuity of $A_r$, note that \eqref{eq:proca j vacuum} implies the “Lorenz-gaugelike” constraint
\begin{equation}
	\label{eq:vacuumLorenz}
	\t\eta{^\alpha^\beta} \t\p{_\alpha} \t A{_\beta} = 0 \,,
\end{equation}
by anti-symmetry of $\t F{_\alpha_\beta}$ and $\t \p{_\alpha} \t j{^\alpha} = 0$.
Since the remaining components have already been shown to be continuous at the interfaces, this can only hold if $A_r$ is continuous as well.

Thus, the boundary condition induced by ideal conductors is
\begin{equation}
	\t A{_\alpha} |_{\p U} = 0 \,.
\end{equation}
Proca's equation \eqref{eq:proca j vacuum} can be equivalently restated as a wave equation
\begin{equation}
	\label{eq:vacuumProcaWave}
	( \Box_\eta - \mu^2 ) \t A{_\alpha} = - \t\eta{_\alpha_\beta} \t j{^\beta} \,,
\end{equation}
keeping in mind that we seek solutions satisfying \eqref{eq:vacuumLorenz}.
Any solution of \eqref{eq:vacuumProcaWave} can be Fourier-decomposed as a sum of fields of the form
\begin{equation}
	\label{eq:Ansatz}
	\t A{_\alpha} = \t{\tilde A}{_\alpha}(x, y) e^{ i ( \beta z - \omega t) } \,,
\end{equation}
leading to the following boundary-value problem for $\t{\tilde A}{_\alpha}$:
\begin{alignat}{2}
	\label{eq:vacuumPDEa}
	-\Delta_\perp \t{\tilde A}{_\alpha}
		& = \xi \t{\tilde A}{_\alpha}
		\quad
		&& \text{in } U \,,
 	\\
	\label{eq:vacuumPDEb}
	\t{\tilde A}{_\alpha}
		&= 0
		&& \text{on } \p U \,,
\end{alignat}
with
\begin{equation}\label{13XII22.5}
 \xi := \omega^2 - \mu^2 - \beta^2
\,,
\end{equation}
and where $\Delta_\perp$ is the two-dimensional Laplacian on the annulus $U$.
We see that $\xi $ belongs to the spectrum of the operator $-\Delta_\perp$ with vanishing Dirichlet data on $\partial U$. It is a standard fact that this spectrum is a discrete subset of $(0,\infty)$; let $\xi_0>0$ denote the smallest eigenvalue. For any non-trivial solution of \eqref{eq:vacuumPDEa}--\eqref{eq:vacuumPDEb} with $\mu>0$ we have
\begin{equation}
	\label{6XII22.2}
	\omega^2 - \beta^2 = \mu^2 + \xi > \xi_0
 	\qquad
 	\implies
 	\qquad
 	\omega^2 > \beta^2 + \xi_0 \,,
\end{equation}
with the last inequality holding independently of $\mu$ as soon as $\mu\ne 0$.
Since the observed TEM modes satisfy
\begin{equation}
	\omega = \pm\beta
\end{equation}
within the precision of measurements, and since $\xi_0>0$, we see that \eqref{6XII22.2} can only be consistent with experiment if $\xi_0$ itself is \emph{not larger} than
the inaccuracy in determining $\omega^2 $ and $\beta^2$. Now, assuming that the inner and outer radii of the conductors are of the same order of magnitude, elementary scaling shows that the smallest eigenvalue $\xi_0$ is of order $r_1^{-2}$.
For example, a numerical approximation for the smallest eigenvalue $\xi_0$ for $r_2 = 2 r_1$ is
\begin{equation}
	\xi_0 \approx 9.75332 \ r_1^{-2} \,,
\end{equation}
which makes clear the incompatibility of \eqref{6XII22.2} with existence of TEM modes.

Our no-go results can be compared with the discussion in \cite{GoldhaberNieto}, where an upper bound on the contribution of the mass of the photon to the perturbation of the Maxwell field has been claimed.
The estimates there are not consistent with our analysis, which shows that some modes disappear completely. (It follows from  \eqref{13XII22.5} that  the surviving ones acquire a phase change of order $\sim \mu^2 L / \beta|_{\mu=0} $, where $L$ is the length of a coaxial cable and $\beta|_{\mu=0}$ is the propagation constant for $\mu=0$.)
The discrepancy can be traced back to the fact that the analysis of \cite{GoldhaberNieto} overlooks the contribution of boundary currents, which are essential in the problem at hand.

\section{Extension to Dielectric Coaxial Cables}

So far we considered wave propagation in vacuum, and the question arises whether the considerations above are affected in the presence of dielectric media.
For this, recall that the Proca equation \eqref{eq:proca j vacuum} is variational, with Lagrangian
\begin{equation}
	L_\text{Proca}
		= - \tfrac{1}{4} \t\eta{^\alpha^\beta} \t\eta{^\gamma^\delta} \t F{_\alpha_\gamma} \t F{_\beta_\delta}
		- \tfrac{1}{2} \mu^2 \t\eta{^\alpha^\beta} \t A{_\alpha} \t A{_\beta}
		+ \t j{^\alpha} \t A{_\alpha}\,.
\end{equation}
To extend the theory to macroscopic electrodynamics in linear isotropic media, we orient ourselves on Maxwell’s equations which, for such media, are derived from the Lagrangian
\begin{equation}
	L_\text{Maxwell}
 		= - \tfrac{1}{4 \permeability} \t\gamma{^\alpha^\beta} \t\gamma{^\gamma^\delta} \t F{_\alpha_\gamma} \t F{_\beta_\delta}
		+ \t j{^\alpha} \t A{_\alpha}\,,
\end{equation}
where $\t\gamma{^\alpha^\beta}$ is Gordon’s optical metric \cite{1923AnP...377..421G}
\begin{equation}
	\t\gamma{^\alpha^\beta}
		= \t\eta{^\alpha^\beta}
		+ (1 - n^2) \t u{^\alpha} \t u{^\beta}\,,
\end{equation}
with $\t u{^\alpha}$ being the four-velocity of the medium and $n = \sqrt{\permeability \permittivity}$ the refractive index expressed in terms of the permittivity $\permittivity$ and permeability $\permeability$.
To pass to a theory of massive photons for macroscopic electrodynamics, we thus consider the Lagrangian
\begin{equation}
	\label{16XI22.1}
	L
		= - \tfrac{1}{4 \permeability} \t\gamma{^\alpha^\beta} \t\gamma{^\gamma^\delta} \t F{_\alpha_\gamma} \t F{_\beta_\delta}
		- \tfrac{1}{2} \mu^2 \t{\mathfrak m}{^\alpha^\beta} \t A{_\alpha} \t A{_\beta}
		+ \t j{^\alpha} \t A{_\alpha}\,,
\end{equation}
where a choice of the “mass matrix” $\t{\mathfrak m}{^\alpha^\beta}$ has to be made. Obvious candidates here are either $\t\eta{^\alpha^\beta}$, or $\t\gamma{^\alpha^\beta}$. We will see that, as before, neither of them allows TEM modes if $\mu\ne 0$. Anticipating, choosing a specific projection matrix as $\t{\mathfrak m}{^\alpha^\beta}$ will allow for such modes, but at the cost of introducing a preferred spatial direction in the equations.

In regions where $ {\permeability} $ is constant, the Lagrangian \eqref{16XI22.1} leads to the following field equations:
\begin{equation}
	\label{eq:proca j dielectric}
	\t\p{_\alpha} (\t\gamma{^\alpha^\rho} \t\gamma{^\beta^\sigma} \t F{_\rho_\sigma})
	- {\permeability} \mu^2 \t{\mathfrak m}{^\alpha^\beta} \t A{_\alpha}
	= - {\permeability} \t j{^\beta} \,.
\end{equation}
In order not to clutter the notation we will from now assume that $\permeability = 1$; alternatively, $\permeability$ can be absorbed in a redefinition of $\mu$ and of $\t j{^\beta}$.

Clearly, there is again a constraint obtained by taking the divergence of \eqref{eq:proca j dielectric},
\begin{equation}
	\label{eq:dielectricLorenz}
	\t\p{_\alpha} ( \t{\mathfrak m}{^\alpha^\beta} \t A{_\beta} ) = 0 \,,
\end{equation}
now “Gordon-Lorenz-gaugelike”.
Additionally, if $\det \t {\mathfrak m}{^\alpha^\beta}\ne 0$, the same arguments as in Section \ref{sec:vacuumProca} show that a) $A_\alpha \equiv 0$ in the conductor region, and b) the boundary condition $ A_\alpha |_{\p U} = 0$ must hold.

Let us start the analysis with the simplest case, which turns out to be ${\mathfrak m}^{\alpha \beta} = \gamma^{\alpha \beta}$. In regions of constant $n$, Equation
\eqref{eq:proca j dielectric} can then be equivalently rewritten as
\begin{equation}
	\label{eq:dielectricProcaWave}
	( \Box_\gamma - \mu^2 ) \t A{_\alpha} = - \t\gamma{_\alpha_\beta } \t j{^\beta} \,.
\end{equation}
Using the Fourier modes \eqref{eq:Ansatz}, one is again led to a spectral problem
\begin{alignat}{2}
	 -\Delta_\perp \t{\tilde A}{_\alpha}
		&= \xi \t{\tilde A}{_\alpha}
		\quad
		&&\text{in } U \,,
	\label{eq:dielectricPDEa}
	\\
	\t{\tilde A}{_\alpha}
	 	&= 0
	 	&&\text{on } \p U \,,
	\label{eq:dielectricPDEb}
\end{alignat}
except that now we have $\xi = n^2 \omega^2 - \mu^2 - \beta^2$.

In the presence of a dielectric, the Maxwell TEM mode propagates with phase velocity $v_p = n^{-1}$, corresponding to $ n\omega = \pm \beta$ at the level of experimental precision.
Since \eqref{eq:dielectricPDEa}--\eqref{eq:dielectricPDEb} is of the same form as \eqref{eq:vacuumPDEa}--\eqref{eq:vacuumPDEb}, an identical argument as in Section~\ref{sec:vacuumProca} leads to the conclusion that $\mu \ne 0$ is incompatible with experiment.

The case of $\t {\mathfrak m}{^\alpha^\beta} = \t\eta{^\alpha^\beta}$ requires more work. The arguments given so far lead to the following boundary-value problem for each Fourier mode \eqref{eq:Ansatz}:
\ptcheck{16XI22, everything so far}
\begin{align}
	-\Delta_\perp \t{\tilde A}{_0}
		& = \xi \t{\tilde A}{_0}\,,
	\label{eq:dielectricPDE2a}
	\\	
	-\Delta_\perp \divergence \tilde A
		 & = \xi \divergence \tilde A
		\,,
	\label{eq:dielectricPDE2b}
	\\
	-\Delta_\perp \t{\tilde A}{_i}
		& = \tilde \xi \t{\tilde A}{_i} + (n^2 - 1) \t\p{_i}
  \divergence \tilde A\,,
	\label{eq:dielectricPDE2c}
\end{align}
in $U$ and
\begin{equation}
	\t{\tilde A}{_\alpha}
	 	 = 0
	 	\ \mbox{on} \ \p U\,,
	\label{eq:dielectricPDE2d}
\end{equation}
with
\begin{align}
	\divergence  \tilde A
		&= \t\p{_x} \t{\tilde A}{_x} + \t\p{_y} \t{\tilde A}{_y} + i \beta \t{\tilde A}{_z}\,,
	\\
	\label{eq:def xi}
	\xi
		&= \omega^2 - \beta^2 - n^{-2} \mu^2\,,
	\\
	\tilde \xi
		&= n^2 \omega^2 - \beta^2 - \mu^2\,.
\end{align}
If $\tilde A_0$ or $\divergence \tilde A$ does not vanish, we obtain a contradiction as before based on the spectral problem associated with \eqref{eq:dielectricPDE2a} or \eqref{eq:dielectricPDE2b}, respectively. The alternative is that $\t {\tilde A}{_0} \equiv 0 \equiv \divergence \tilde A$, in which case \eqref{eq:dielectricPDE2c} becomes
\begin{equation}
	- \Delta_\perp \tilde{A}_{i}
	= \tilde \xi \tilde{A}_{i} \,,
\end{equation}
and a contradiction with existence of TEM modes is obtained from the inequalities
\begin{equation}
	n^2 \omega^2 - \beta^2 = \mu^2+ \tilde \xi > \xi_0
 	\qquad
 	\implies
 	\qquad
 	n^2 \omega^2 > \beta^2 +\xi_0 \,,
\end{equation}
with $\xi_0$ as in \eqref{6XII22.2}.

Summarizing, we have shown that the Proca field equations are not compatible with TEM modes in a coaxial cable, whether in vacuum or in dielectric media, contradicting the fact that TEM modes constitute the main propagating mode in coaxial cables. Thus, neither the Proca equation, nor its variation involving the Gordon metric, correctly describe photons with small non-zero mass in this setting.

Our analysis provides yet another piece of evidence that photons are massless. The question then arises, whether there exist an equation which sidesteps the problem raised above. The requirement of Lorentz covariance allows for modifications of the Proca Lagrangian which involve only the Lorentz metric and the four-velocity $\t u{^\alpha}$ of the medium. For reasons that will become clear shortly, we also need a unit vector field $\t\ell{^\alpha}$, orthogonal to $\t u{^\alpha}$ and directed along the axis of symmetry of the medium, and consider the Lagrangian \eqref{16XI22.1} with $\t{\mathfrak m}{^\alpha^\beta}$ given by
\begin{equation}
	\label{27II22.5}
	\t\matM{^\alpha^\beta }
		= \t \eta{^\alpha^\beta }
		+ p \t u{^\alpha} \t u{^\beta}
		- q \t \ell{^\alpha} \t \ell{^\beta} \,,
\end{equation}
for some constants $p$ and $q$ to be determined.
This is the most general form of $\t{\mathfrak m}{^\alpha^\beta}$ which is compatible with Lorentz invariance, time-reversal symmetry, the existence of a preferred four-velocity determined by the rest frame of the medium, and the existence of a preferred spacelike direction determined by the geometry of the medium.

In regions where $\t\p{_\sigma} \t u{^\alpha}$, $\t\p{_\sigma} \t \ell{^\alpha}$ abd $\t\p{_\sigma} n$ vanish, the resulting field equations read
\begin{equation}
	\label{eq:ModifiedProcaFieldEquations}
	\t \gamma{^\alpha^\beta} \t\p{_\alpha} \t F{_\beta_\gamma}
	- \mu^2 \t A{_\alpha} \t\matM{^\alpha^\beta} \t\gamma{_\beta_\gamma} = 0 \,.
\end{equation}
A repetition of the arguments above shows that, if $\mu \neq 0$, the vanishing of $F_{\alpha\beta}$ in the conducting region implies $\red{\t A{_\alpha}} \t\matM{^\alpha^\beta} \t\gamma{_\beta_\gamma} = 0$ there.
The contradictions reached above can only be avoided if we can find field configurations for which $\t A{_\alpha} \not \equiv 0$ in the conducting region. This will be the case if and only if the matrix
\begin{equation}
	\t \matM{^\alpha^\beta} \t \gamma{_\beta_\gamma}
	= \t*\delta{^\alpha_\gamma}
	+ \left( 1 - \frac{ 1 - p}{ n^2 } \right) \t u{^\alpha} \t u{^\beta} \t\eta{_\beta_\gamma}
	- q \, \t \ell{^\alpha} \t\ell{^\beta} \t\eta{_\beta_\gamma}
\end{equation}
is singular.
The vanishing of the determinant of $\t \matM{^\alpha^\beta} \t\gamma{_\beta_\gamma}$ is equivalent to $( p - 1 ) ( q - 1 ) = 0$.
We show in an Appendix that inconsistencies arise again unless $p = 1 = q$, which leads to an equation which \emph{does} now admit modes approximating the observed TEM ones:
\begin{equation}
	\t\gamma{^\alpha^\beta} \t\p{_\alpha} \t F{_\beta_i}
	= 0
	\quad \text{with} \quad i \in \{0,3\}\,,
\end{equation}
and
\begin{equation}
	\t\gamma{^\alpha^\beta} \t\p{_\alpha} \t F{_\beta_j} - \mu^2 \t A{_j} = 0
	\quad \text{with} \quad j \in \{1,2\}\,.
\end{equation}
In such a theory there is a residual gauge freedom $A_\mu\to A_\mu + \partial_\mu \lambda$, where $\lambda$ is an arbitrary function of $t$ and $z$, and only the transverse components of the field acquire a mass.

But, none of the equations above are compatible with experimental evidence if $\mu\ne 0$, no matter how small, without introducing extra structures in the equation.

One might wonder whether an alternative solution could be found by introducing in \eqref{16XI22.1}
a new optical metric
\begin{equation}
	\t\gamma{^\alpha^\beta }
		= \t \eta{^\alpha^\beta }
		+ \redsigma \t u{^\alpha} \t u{^\beta}
		+ \redtau \t \ell{^\alpha} \t \ell{^\beta} \,,
\end{equation}
with constants $\sigma$ and $\tau$ restricted to a range so that $\t \gamma{^\alpha^\beta}$ is Lorentzian, to guarantee a well-posed Cauchy problem. But a linear rescaling of the coordinates together with obvious field redefinitions reduces the resulting equations to the ones already analyzed above, leading to the same incompatibilities with experiment.

\section{Conclusions}

We have shown that a non-zero photon mass $\mu$ in the Proca equation implies interface conditions for the electromagnetic field which deviate strongly from those of Maxwell’s equations, irrespective of the magnitude of $\mu$.
Combined with the assumptions listed in the Introduction, Proca’s equation thus predicts an absence of TEM modes in coaxial cables, which are both predicted to exist by Maxwell’s equations and observed in experiments.
From this, we conclude that the following statements are incompatible:
(i) Proca’s equation holds,
(ii) charge conservation holds,
(iii) the electromagnetic field has an everywhere well-defined energy-momentum density,
(iv) ideal conductors are admissible idealizations in electrodynamics.

It was argued in~\cite{GoldhaberNieto,1971Kroll} that the dispersion relation of the TEM mode is insensitive to the photon mass $\mu$.
Our results here prove rigorously instead that the observed existence of TEM modes is \emph{incompatible} with the Proca equation.

\appendix

\section*{Appendix}
\label{app6XII22}

In this appendix we show that TE modes, a special case of which are the TEM modes, are incompatible with \eqref{eq:ModifiedProcaFieldEquations} unless $p = q = 1$. For this, recall that the TE modes of Maxwell’s equations satisfy
$\t E{_z}= - i(\beta \t{\tilde A}{_0} + \omega \t{\tilde A}{_z}) e^{i(\beta z - \omega t)} \equiv 0$ in $U$.
Hence, a solution of the Proca equation which approximates the observed Maxwell TEM solutions has to satisfy
\begin{equation}
	\label{eq:TE}
	\beta \t{\tilde A}{_0} + \omega \t{\tilde A}{_z} = f \,,
\end{equation}
where $f$ is a function of $\mu$ and all coordinates which is not larger than the experimental precision, say $f=O(\epsilon)$.
Using $f$, the system \eqref{eq:ModifiedProcaFieldEquations} can be written as
\begin{align}
	(-\Delta_\perp-\hat \xi) \t{\tilde A}{_0}
		&= i q \t\p{_z} f\,,
	\label{7XII22.11}
	\\
	(-\Delta_\perp-\check \xi) \t{\tilde A}{_z}
		&= - i (1 - p - n^2 ) \t\p{_0} f\,,
	\label{7XII22.12}
	\\
	(-\Delta_\perp - \xi) \t{\tilde A}{_x}
		&= - i \p_{x} (q \beta \t{\tilde A}{_z} - (1 - p - n^2) \omega \t{\tilde A}{_0})\,,
	\label{5XII22.1}
	\\
	(-\Delta_\perp - \xi) \t{\tilde A}{_y}
		&= - i \p_{y} (q \beta \t{\tilde A}{_z} - (1 - p - n^2) \omega \t{\tilde A}{_0})\,,
	\label{5XII22.2}
\end{align}
where $\xi$ is as in \eqref{eq:def xi} and
\begin{align}
	\hat \xi
		&= \omega^2 (1-p) - \beta^2 (1- q) - \mu^2 n^{-2} (1 - p)\,,
	\\
	\check \xi
		&= \omega^2 (1-p) - \beta^2 (1- q) - \mu^2 (1 - q) \,.
\end{align}
\ptcheck{6XII22}

Take now $p = 1$ and $q \neq 1$. With this mass matrix $\t{\mathfrak m}{^\alpha^\beta}$ we still need $\t{\tilde A}{_i} |_{\p U} = 0$ but now $\t{\tilde A}{_0} |_{\p U}$ can be arbitrary. Then \eqref{7XII22.12} implies that $\t{\tilde A}{_z} = O(\epsilon)$ unless $\check \xi$ is an eigenvalue. In this last case we obtain a contradiction as in the main body of this paper, thus it must be that $\t{\tilde A}{_z} = O(\epsilon)$ holds. But then \eqref{eq:TE} shows that $\t{\tilde A}{_0} = O(\epsilon)$ as well. It follows that the source terms in \eqref{5XII22.1}-\eqref{5XII22.2} are $O(\epsilon)$, and the whole solution will be $O(\epsilon)$ unless $\xi$ is an eigenvalue. An inconsistency with experiment is obtained by \eqref{6XII22.2}.

The treatment for $q = 1$ but $p \neq 1$ is analogous.
This establishes our claim.

\section*{Acknowledgements}
Research supported in part by the Austrian Science Fund (FWF), Project P34274 and Grant TAI 483-N, as well as the Vienna University Research Platform TURIS. TBM is a recipient of a DOC Fellowship of the Austrian Academy of Sciences at the Faculty of Physics of the University of Vienna and acknowledges partial support from the Vienna Doctoral School in Physics (VDSP).

\providecommand{\bysame}{\leavevmode\hbox to3em{\hrulefill}\thinspace}
\providecommand{\MR}{\relax\ifhmode\unskip\space\fi MR }
\providecommand{\MRhref}[2]{%
  \href{http://www.ams.org/mathscinet-getitem?mr=#1}{#2}
}
\providecommand{\href}[2]{#2}

\end{document}